\documentclass[12pt]{article}

\usepackage{amsmath,amssymb,array,calc,rotating,epsfig,psfrag, amscd, graphicx,color}
\numberwithin{equation}{section}
\usepackage[left=2cm,top=25mm,bottom=25mm,right=2cm,nohead]{geometry}

\usepackage[all]{xy}

\usepackage[
      colorlinks=true,
       linkcolor=darkblue,  
      urlcolor=blue,    
      filecolor=blue,     
      citecolor=red,
      pdfstartview=FitV,
       bookmarksopen=true    
      ]{hyperref}
      
\newcommand{\nc}{\newcommand}

\definecolor{cardinal}{rgb}{0.6,0,0}
\definecolor{darkgreen}{rgb}{0,0.5,0}
\definecolor{golden}{rgb}{0.92, 0.7, 0}
\definecolor{midnight}{rgb}{0, 0, 0.5}
\definecolor{darkblue}{rgb}{0.2, 0, 0.8}

\nc{\ra}{\rightarrow} 
\nc{\lra}{\leftrightarrow} 
\nc{\Ra}{\Rightarrow} 
\nc{\LRa}{\Leftightarrow} 
\nc{\blp}{{\big (}}
\nc{\brp}{{\big )}}
\nc{\Blp}{{\Big (}}
\nc{\Brp}{{\Big )}}
\nc{\bglp}{{\bigg (}}
\nc{\bgrp}{{\bigg )}}
\nc{\Bglp}{{\Bigg (}}
\nc{\Bgrp}{{\Bigg )}}
\nc{\slb}{{\rm [}}
\nc{\srb}{{\rm ]}}
\nc{\bslb}{{\rm \big [}}
\nc{\bsrb}{{\rm \big ]}}
\nc{\Bslb}{{\rm \Big [}}
\nc{\Bsrb}{{\rm \Big ]}}

\def\al{\alpha}

\def\eps{\epsilon}
\nc{\veps}{\varepsilon}
\def\gam{\gamma}

\def\lam{\lambda}
\def\om{\omega}

\nc{\vphi}{\varphi}
\def\tha{\theta}

\def\sig{\sigma}

\def\Gam{\Gamma}

\def\Lam{\Lambda}
\def\Om{\Omega}
\def\Sig{\Sigma}

\def\coeff#1#2{\relax{\textstyle {#1 \over #2}}\displaystyle}

\def\fp{\frak{p}}
\def\fq{\frak{q}}

\nc{\myvspace}{\rule[-1em]{0pt}{2.5em}}
\nc{\bea}{\begin{eqnarray}}
\nc{\eea}{\end{eqnarray}}
\nc{\be}{\begin{equation}}
\nc{\ee}{\end{equation}}
\nc{\barr}{\begin{array}}
\nc{\earr}{\end{array}}
\nc{\cA}{{\cal A}}
\nc{\cB}{ \cal B}

\def\cD{{\cal D}}

\nc{\cF}{{\cal F}}
\nc{\cG}{{\cal G}}

\def\cI{{\cal I}}

\nc{\cL}{{\cal L}}
\nc{\cM}{{\cal M}}

\def\cN{{\cal N}}
\def\cNbar{\ol{\cal N}}

\def\cP{{\cal P}}
\nc{\cQ}{{\cal Q}}
\nc{\cR}{{\cal R}}
\def\cS{{\cal S}}

\def\cV{{\cal V}}
\def\cV{{\cal V}}
\def\cW{{\cal W}}

\def\cZ{{\cal Z}}
\nc{\cQd}{\cQ^{\dagger}}
\nc{\cRd}{\cR^{\dagger}}
\nc{\BB}{{\mathbb B}}
\nc{\CC}{{\mathbb C}}
\nc{\DD}{{\mathbb D}}
\nc{\EE}{{\mathbb E}}
\nc{\FF}{{\mathbb F}}
\nc{\GG}{{\mathbb G}}
\nc{\HH}{{\mathbb H}}
\nc{\JJ}{{\mathbb J}}
\nc{\MM}{{\mathbb M}}
\nc{\RR}{{\mathbb R}}
\nc{\PP}{{\mathbb P}}
\nc{\QQ}{{\mathbb Q}}
\nc{\UU}{{\mathbb U}}
\nc{\ZZ}{{\mathbb Z}}
\nc{\calone}{{\mathbb 1}}
\nc{\half}{\coeff{1}{2}}
\nc{\quarter}{\coeff{1}{4}}
\nc{\del}{\partial}

\nc{\delbar}{\bar\partial}
\nc{\thalf}{\frac{t}{2}}
\nc{\Spin}{\operatorname{Spin}}
\nc{\SO}{\operatorname{SO}}

\nc{\Sp}{{\rm Sp}}
\nc{\com}[2]{{ \left[ #1, #2 \right] }}
\nc{\acom}[2]{{ \left\{ #1, #2 \right\} }}
\nc{\rr}{\rightarrow}
\nc{\p}{\partial}
\nc{\LT}{{\LL_\T}}
\nc{\Tr}{{\rm Tr}}
\nc{\tr}{{\rm tr}}
\nc{\Adag}{A^{\dagger}}
\nc{\AdagI}{A^{\dagger I}}
\nc{\AdagJ}{A^{\dagger J}}
\nc{\AdagK}{A^{\dagger K}}
\nc{\AdagL}{A^{\dagger L}}
\nc{\AdagM}{A^{\dagger M}}
\nc{\Bdag}{B^{\dagger}}
\nc{\BdagI}{B^{\dagger}_I}
\nc{\BdagJ}{B^{\dagger}_J}
\nc{\BdagK}{B^{\dagger}_K}
\nc{\BdagL}{B^{\dagger}_L}
\nc{\BdagM}{B^{\dagger}_M}
\nc{\Cdag}{C^{\dagger}}
\nc{\CdagI}{C^{\dagger I}}
\nc{\CdagJ}{C^{\dagger J}}
\nc{\CdagK}{C^{\dagger K}}
\nc{\Ddag}{D^{\dagger}}
\nc{\DdagI}{D^{\dagger I}}
\nc{\DdagJ}{D^{\dagger J}}
\nc{\DdagK}{D^{\dagger K}}
\nc{\ttha}{\tilde{\theta}}
\nc{\ttau}{\tilde{\tau}}
\nc{\tTha}{\tilde{\Theta}}
\nc{\tphi}{\tilde{\phi}}
\nc{\tsig}{\tilde{\sig}}
\nc{\tom}{\widetilde{\om}}
\nc{\tOm}{\widetilde{\Om}}
\nc{\tlam}{\widetilde{\lam}}
\nc{\tLam}{\tilde{\Lam}}
\nc{\tSig}{\widetilde{\Sig}}
\nc{\tPhi}{\tilde{\Phi}}
\nc{\tPhibar}{\ol{\tPhi}}
\nc{\tPi}{\widetilde{\Pi}}
\nc{\tpsi}{\widetilde{\psi}}
\nc{\tPsi}{\tilde{\Psi}}
\nc{\tgam}{\widetilde{\gam}}
\nc{\tGam}{\widetilde{\Gam}}
\nc{\tzeta}{\tilde{\zeta}}
\nc{\tZeta}{\tilde{\Zeta}}
\nc{\teta}{\widetilde{\eta}}
\nc{\teps}{\tilde{\eps}}
\nc{\tveps}{\tilde{\veps}}
\nc{\tEta}{\tilde{\Eta}}
\nc{\tchi}{\tilde{\chi}}
\nc{\tChi}{\tilde{\Chi}}
\nc{\txi}{\tilde{\xi}}
\nc{\tXi}{\widetilde{\Xi}}
\nc{\tnu}{\tilde{\nu}}
\nc{\tmu}{\tilde{\mu}}

\nc{\tb}{\tilde b}
\nc{\tc}{\tilde c}
\nc{\te}{\tilde e}
\nc{\tf}{\tilde f}
\nc{\tg}{\tilde g}
\nc{\ti}{\tilde i}
\nc{\tj}{\tilde j}
\nc{\tk}{\tilde k}
\nc{\tl}{\tilde l}
\nc{\tm}{\tilde m}
\nc{\tn}{\tilde n}
\nc{\tp}{\tilde{p}}
\nc{\tq}{\widetilde{q}}
\nc{\ts}{{\tilde s}}
\nc{\tu}{{\tilde u}}
\nc{\tv}{{\tilde v}}
\nc{\tw}{{\tilde w}}
\nc{\tx}{{\tilde x}}
\nc{\ty}{{\tilde y}}
\nc{\tz}{\tilde z}
\nc{\tA}{{\widetilde A}}
\nc{\tAbar}{{\ol \tA}}
\nc{\tB}{{\widetilde B}}
\nc{\tC}{{\widetilde C}}
\nc{\tD}{{\widetilde D}}
\nc{\tE}{{\widetilde E}}
\nc{\tF}{{\widetilde F}}
\nc{\tG}{{\widetilde G}}
\nc{\tH}{{\widetilde H}}
\nc{\tJ}{{\widetilde J}}
\nc{\tJbar}{{\ol {\tilde J}}}
\nc{\tK}{{\widetilde K}}
\nc{\tL}{{\widetilde L}}
\nc{\tcL}{{\widetilde \cL}}
\nc{\tM}{{\widetilde M}}
\nc{\tN}{{\widetilde N}}
\nc{\tcN}{{\widetilde \cN}}
\nc{\tP}{{\widetilde P}}
\nc{\tQ}{{\widetilde Q}}
\nc{\tR}{{\widetilde R}}
\nc{\tS}{\widetilde{S}}
\nc{\tT}{\widetilde{T}}
\nc{\tU}{\widetilde{U}}
\nc{\tV}{\widetilde{V}}
\nc{\tW}{\widetilde{W}}
\nc{\tcF}{\widetilde{{\cal F}}}
\nc{\tX}{\widetilde{X}}
\nc{\tY}{\widetilde{Y}}
\nc{\tcZ}{\tilde{\cZ}}
\nc{\tcZbar}{\ol{\tcZ}}

\nc{\ha}{\hat a}
\nc{\hb}{\hat b}
\nc{\hc}{\widehat c}
\nc{\hd}{\widehat d}
\nc{\he}{\widehat e}
\nc{\hf}{\widehat f}
\nc{\hg}{\widehat g}
\nc{\hh}{\widehat h}
\nc{\hm}{\widehat m}
\nc{\hn}{\widehat n}
\nc{\hp}{\widehat p}
\nc{\hr}{\widehat r}
\nc{\hs}{\widehat s}
\nc{\hv}{\widehat v}
\nc{\hw}{\widehat w}
\nc{\hx}{\widehat x}
\nc{\hy}{\widehat y}
\nc{\hz}{\widehat z}
\nc{\zhat}{\hat z}
\nc{\hA}{\widehat{A}}
\nc{\hB}{\widehat{B}}
\nc{\hC}{\widehat{C}}
\nc{\hD}{\widehat{D}}
\nc{\hE}{\widehat{E}}
\nc{\hF}{\widehat{F}}
\nc{\hcF}{\widehat{\cF}}
\nc{\hG}{\widehat{G}}
\nc{\hH}{\widehat{H}}
\nc{\hJ}{\widehat{J}}
\nc{\hK}{\widehat{K}}
\nc{\hL}{\widehat{L}}
\nc{\hcL}{\widehat{\cL}}
\nc{\hM}{\widehat M}
\nc{\hcM}{\widehat{\cM}}
\nc{\hN}{\widehat{N}}
\nc{\hO}{\widehat{O}}
\nc{\hP}{\widehat{P}}
\nc{\hQ}{\widehat{Q}}
\nc{\hcR}{\widehat{\cR}}
\nc{\hR}{\widehat{R}}
\nc{\hS}{\widehat{S}}
\nc{\hcS}{\widehat{\cS}}
\nc{\hT}{\widehat{T}}
\nc{\hU}{\widehat{U}}
\nc{\hV}{\widehat V}
\nc{\hcV}{\widehat \cV}
\nc{\hX}{\widehat X}
\nc{\hcZ}{\widehat \cZ}
\nc{\hcZbar}{\ol{\widehat \cZ}}

\nc{\heta}{\widehat{\eta}}
\nc{\hal}{\widehat \alpha}
\nc{\hphi}{\widehat{\phi}}
\nc{\hkap}{\hat{\kappa}}
\nc{\hchi}{\widehat{\chi}}
\nc{\hpsi}{\widehat{\psi}}
\nc{\hgam}{\widehat{\gam}}
\nc{\hPhi}{\hat{\Phi}}
\nc{\hPsi}{\hat{\Psi}}
\nc{\hGam}{\hat{\Gam}}
\nc{\omhat}{\widehat{\om}}
\nc{\htha}{\hat{\tha}}
\nc{\hrho}{\widehat{\rho}}
\nc{\hdel}{\widehat{\del}}

\nc{\w}{\wedge}

\nc{\vb}{\vec b}
\nc{\vc}{\vec c}
\nc{\vd}{\vec d}
\nc{\ve}{\vec e}
\nc{\vf}{\vec f}
\nc{\vg}{\vec g}
\nc{\vh}{\vec h}
\nc{\vp}{\vec p}
\nc{\vq}{\vec q}
\nc{\vr}{\vec r}
\nc{\vs}{\vec s}
\nc{\vv}{\vec v}
\nc{\vw}{\vec w}
\nc{\vx}{\vec x}
\nc{\vy}{\vec y}
\nc{\vz}{\vec z}

\nc{\vB}{\vec B}
\nc{\vC}{\vec C}
\nc{\vD}{\vec D}
\nc{\vE}{\vec E}
\nc{\vF}{\vec F}
\nc{\vG}{\vec G}
\nc{\vH}{\vec H}
\nc{\vP}{\vec P}
\nc{\vQ}{\vec Q}
\nc{\vR}{\vec R}
\nc{\vS}{\vec S}
\nc{\vV}{\vec V}
\nc{\vW}{\vec W}
\nc{\vX}{\vec X}
\nc{\vY}{\vec Y}
\nc{\vZ}{\vec Z}

\nc{\ol}{\overline}
\nc{\abar}{\ol{a}}
\nc{\bbar}{\ol{b}}
\nc{\cbar}{\ol{c}}
\nc{\dbar}{\ol{d}}
\nc{\ebar}{\ol{e}}
\nc{\fbar}{\ol{f}}
\nc{\ibar}{\ol{\imath}}
\nc{\jbar}{\ol{\jmath}}
\nc{\kbar}{\ol{k}}
\nc{\lbar}{\ol{l}}
\nc{\mbar}{\ol{m}}
\nc{\nbar}{\ol{n}}
\nc{\pbar}{\ol{p}}
\nc{\qbar}{\ol{q}}
\nc{\rbar}{\ol{r}}
\nc{\sbar}{\ol{s}}
\nc{\ubar}{\ol{u}}
\nc{\vbar}{\ol{v}}
\nc{\wbar}{\ol{w}}
\nc{\xbar}{\ol{x}}
\nc{\ybar}{\ol{y}}
\nc{\zbar}{\ol{z}}

\nc{\Abar}{\ol{A}}
\nc{\Bbar}{\ol{B}}
\nc{\Cbar}{\ol{C}}
\nc{\Dbar}{\ol{D}}
\nc{\Ebar}{\ol{E}}
\nc{\Fbar}{\ol{F}}
\nc{\Jbar}{\ol{J}}
\nc{\Kbar}{\ol{K}}
\nc{\Lbar}{\ol{L}}
\nc{\cLbar}{\ol{\cL}}
\nc{\Mbar}{\ol{M}}
\nc{\Nbar}{\ol{N}}
\nc{\Pbar}{\ol{P}}
\nc{\Qbar}{\ol{Q}}
\nc{\Rbar}{\ol{R}}
\nc{\Sbar}{\ol{S}}
\nc{\Tbar}{\ol{T}}
\nc{\Ubar}{\ol{U}}
\nc{\Vbar}{\ol{V}}
\nc{\cVbar}{\ol{\cV}}
\nc{\Wbar}{\ol{W}}
\nc{\cWbar}{\ol{\cW}}
\nc{\Xbar}{{\overline X}}
\nc{\Ybar}{{\overline Y}}
\nc{\Zbar}{{\overline Z}}
\nc{\cZbar}{{\overline \cZ}}

\nc{\epsbar}{\ol{\epsilon}}
\nc{\lambar}{\ol{\lambda}}
\nc{\kapbar}{\ol{\kappa}}
\nc{\zetabar}{\ol{\zeta}}
\nc{\Zetabar}{\ol{\Zeta}}
\nc{\taubar}{\ol{\tau}}
\nc{\Taubar}{\ol{\Tau}}
\nc{\psibar}{\ol{\psi}}
\nc{\Psibar}{\ol{\Psi}}
\nc{\tpsibar}{\ol{\tpsi}}
\nc{\tPsibar}{\ol{\tPsi}}
\nc{\phibar}{\ol{\phi}}
\nc{\Phibar}{\ol{\Phi}}
\nc{\chibar}{\ol{\chi}}
\nc{\mubar}{\ol{\mu}}
\nc{\nubar}{\ol{\nu}}
\nc{\rhobar}{\ol{\rho}}
\nc{\ombar}{\ol{\om}}
\nc{\Ombar}{\ol{\Om}}
\nc{\Deltabar}{\ol{\Delta}}
\nc{\Thetabar}{\ol{\Theta}}
\nc{\xibar}{\ol{\xi}}
\nc{\Xibar}{\ol{\Xi}}

\nc{\Dthbar}{\ol{\rm D3}}

\nc{\gdot}{\dot{g}}
\nc{\pdot}{\dot{p}}
\nc{\qdot}{\dot{q}}
\nc{\rdot}{\dot{r}}
\nc{\sdot}{\dot{s}}
\nc{\tdot}{\dot{t}}
\nc{\udot}{\dot{u}}
\nc{\vdot}{\dot{v}}
\nc{\wdot}{\dot{w}}
\nc{\xdot}{\dot{x}}
\nc{\xddot}{\ddot{x}}
\nc{\ydot}{\dot{y}}
\nc{\zdot}{\dot{z}}
\nc{\yddot}{\ddot{y}}

\nc{\Udot}{\dot{U}}
\nc{\Vdot}{\dot{V}}
\nc{\Wdot}{\dot{W}}

\nc{\taudot}{\dot{\tau}}
\nc{\phidot}{\dot{\phi}}
\nc{\psidot}{\dot{\psi}}
\nc{\sinp}{s_{\phi}}
\nc{\cosp}{c_{\phi}}
\nc{\tanp}{t_{\phi}}
\nc{\spone}{s_{\phi_1}}
\nc{\cpone}{c_{\phi_1}}
\nc{\tpone}{t_{\phi_1}}
\nc{\sptwo}{s_{\phi_2}}
\nc{\cptwo}{c_{\phi_2}}
\nc{\tptwo}{t_{\phi_2}}
\nc{\spth}{s_{\phi_3}}
\nc{\cpth}{c_{\phi_3}}
\nc{\tpth}{t_{\phi_3}}
\nc{\calp}{c_{\al}}
\nc{\salp}{s_{\al}}

\nc{\csch}{{\rm csch}}
\nc{\sech}{{\rm sech}}

\nc{\cothzlami}{\coth(z-\lam_i)}
\nc{\coshzlami}{\cosh(z-\lam_i)}
\nc{\sinhzlami}{\sinh(z-\lam_i)}

\nc{\cothzlamj}{\coth(z-\lam_j)}
\nc{\coshzlamj}{\cosh(z-\lam_j)}
\nc{\sinhzlamj}{\sinh(z-\lam_j)}

\nc{\cothlamij}{\coth(\lam_i-\lam_j)}
\nc{\coshlamij}{\cosh(\lam_i-\lam_j)}
\nc{\sinhlamij}{\sinh(\lam_i-\lam_j)}

\nc{\bah}{{\mathbf {\hat{A}}}}
\nc{\bX}{{\mathbf X}}
\nc{\ba}{{\bf a}}
\nc{\bb}{{\bf b}}
\nc{\bc}{{\bf c}}
\nc{\bd}{{\bf d}}
\nc{\bg}{{\bf g}}
\nc{\bk}{{\bf k}}
\nc{\bl}{{\bf l}}
\nc{\bm}{{\bf m}}
\nc{\bn}{{\bf n}}
\nc{\bo}{{\bf o}}
\nc{\bp}{{\bf p}}
\nc{\bq}{{\bf q}}
\nc{\br}{{\bf r}}
\nc{\bs}{{\bf s}}
\nc{\bt}{{\bf t}}
\nc{\bu}{{\bf u}}
\nc{\bv}{{\bf v}}
\nc{\bw}{{\bf w}}
\nc{\bx}{{\bf x}}
\nc{\by}{{\bf y}}
\nc{\bz}{{\bf z}}
\nc{\bom}{{\bf \om}}
\nc{\bombar}{{\mathbf \ombar}}
\nc{\bPhi}{{\bf \Phi}}

\nc{\rma}{{\rm a}}
\nc{\rmb}{{\rm b}}
\nc{\rmc}{{\rm c}}
\nc{\rmd}{{\rm d}}
\nc{\rmg}{{\rm g}}
\nc{\rk}{{\rm k}}
\nc{\rml}{{\rm l}}
\nc{\rmm}{{\rm m}}
\nc{\rmn}{{\rm n}}
\nc{\rmo}{{\rm o}}
\nc{\rmp}{{\rm p}}
\nc{\rmq}{{\rm q}}
\nc{\rmr}{{\rm r}}
\nc{\rms}{{\rm s}}
\nc{\rmt}{{\rm t}}
\nc{\rmu}{{\rm u}}
\nc{\rmv}{{\rm v}}
\nc{\rmw}{{\rm w}}
\nc{\rmx}{{\rm x}}
\nc{\rmy}{{\rm y}}
\nc{\rmz}{{\rm z}}

\nc{\dal}{\dot{\al}}
\nc{\thadot}{\dot{\tha}}
\nc{\thab}{\bar{\theta}}
\nc{\thal}{\theta^{\al}}
\nc{\thdal}{\bar{\theta}^{\dal}}

\nc{\thsigthm}{\tha \sigma^m \thab}
\nc{\thsigthn}{\tha \sigma^n \thab}

\nc{\Dal}{D_{\al}}
\nc{\Ddal}{\bar{D}_{\dal}}
\nc{\CDal}{{\cal D}_{\al}}
\nc{\CDdal}{\bar{\cal D}_{\dal}}

\nc{\eq}[1]{(\ref{#1})}
\nc{\non}{\nonumber}
\nc{\Fzero}{F_{(0)}}
\nc{\Ftwo}{F_{(2)}}
\nc{\Ffour}{F_{(4)}}
\nc{\Fone}{F_{(1)}}
\nc{\Fthree}{F_{(3)}}
\nc{\Ffive}{F_{(5)}}
\nc{\Fn}{F_{(n)}}
\nc{\Fp}{F_{(p)}}

\nc{\tFzero}{\tF_{(0)}}
\nc{\tFtwo}{\tF_{(2)}}
\nc{\tFfour}{\tF_{(4)}}
\nc{\tFone}{\tF_{(1)}}
\nc{\tFthree}{\tF_{(3)}}
\nc{\tFfive}{\tF_{(5)}}
\nc{\tFn}{\tF_{(n)}}
\nc{\tFp}{\tF_{(p)}}

\nc{\Czero}{C_{(0)}}
\nc{\Ctwo}{C_{(2)}}
\nc{\Cfour}{C_{(4)}}
\nc{\Cone}{C_{(1)}}
\nc{\Cthree}{C_{(3)}}
\nc{\Cfive}{C_{(5)}}
\nc{\Cn}{C_{(n)}}

\nc{\equ}{{\rm eq}}
\def\Im{{\rm Im \hspace{0.5mm} }}

\nc{\vol}{{\rm vol}}
\nc{\Ainf}{A_{\infty}}
\nc{\End}{{\rm End}}
\nc{\Ext}{{\rm Ext}}
\nc{\IIB}{{\rm IIB}}
\nc{\Ad}{{\rm Ad}}
\nc{\IIA}{{\rm IIA}}
\nc{\AdS}{{\rm AdS}}
\nc{\CFT}{{\rm CFT}}
\nc{\diag}{{\rm diag}}
\nc{\Log}{{\rm Log}}
\nc{\Dslash}{\ensuremath \raisebox{0.025cm}{\slash}\hspace{-0.32cm} D}
\nc{\cDslash}{\ensuremath \raisebox{0.025cm}{\slash}\hspace{-0.32cm} \cD}
\nc{\omslash}{\om\!\!\!/}
\nc{\no}{\!:\!\!}
\nc{\ointdz}{\oint\frac{dz}{2\pi i}}
\nc{\ointdzone}{\oint\frac{dz_1}{2\pi i}}
\nc{\ointdztwo}{\oint\frac{dz_2}{2\pi i}}
\nc{\ointdzb}{\oint\frac{d\zbar}{2\pi i}}
\nc{\ointdzbone}{\oint\frac{d\zbar_1}{2\pi i}}
\nc{\ointdzbtwo}{\oint\frac{d\zbar_2}{2\pi i}}
\nc{\dz}{\frac{dz}{2\pi i}}
\nc{\dzb}{\frac{d\zbar}{2\pi i}}
\nc{\bpm}{\begin{pmatrix}}
\nc{\epm}{\end{pmatrix}}
 \nc{\bitem}{\begin{itemize}}
 \nc{\eitem}{\end{itemize}}
 \nc{\exercise}{\vskip 2mm \noindent {\bf Exercise:}}
 \nc{\definition}{\vskip 2mm \noindent {\bf Definition:}}
 
\begin{document}
%%%%%%%%%%%%%%%%%%%%%%%%%%%%%%%%%%%%%%%%%%%%%%%%%%%%%%%%%%%%%%%%%
%%%%%%%%%%%%%%%%%%%%%%%%%%%%%%%%%%%%%%%%%%%%%

\vspace{0.5cm}
\begin{center}
\baselineskip=13pt {\LARGE \bf{BPS Black Hole Horizons in  \\ N=2 Gauged Supergravity\\}}
\vskip 20mm
Nick Halmagyi\\ 
\vskip0.5cm
Laboratoire de Physique Th\'eorique et Hautes Energies,\\
Universit\'e Pierre et Marie Curie, CNRS UMR 7589, \\
F-75252 Paris Cedex 05, France\\
\vskip0.5cm
halmagyi@lpthe.jussieu.fr \\ 
\end{center}

\vskip 15mm
\begin{abstract}
We study static BPS black hole horizons in four dimensional $\cN=2$ gauged supergravity coupled to $n_v$-vector multiplets and with an arbitrary cubic prepotential. We work in a symplectically covariant formalism which allows for both electric and magnetic gauging parameters as well as dyonic background charges and obtain the general solution to the BPS equations for horizons of the form $AdS_2\times \Sigma_g$. In particular this means we solve for the scalar fields as well as the metric of these black holes as a function of the gauging parameters and background charges. When the special K\"ahler manifold is a symmetric space, our solution is completely explicit and the entropy is related to the familiar quartic invariant. For more general models our solution is implicit up to a set of holomorphic quadratic equations. For particular models which have known embeddings in M-theory, we derive new horizon geometries with dyonic charges and numerically construct black hole solutions. These  correspond to M2-branes wrapped on a Riemann surface in a local Calabi-Yau five-fold with internal spin. 
\end{abstract}

\vskip 15mm
\begin{center}
{\it Dedicated to the loving memory of Sue Halmagyi}
\end{center}

\newpage

%%%%%%%%%%%%%%%%%%%%%%%%%%%%%%%%%%%%%%%%%%%%%%%%%%%%
%%%%%%%%%%%%%%%%%%%%%%%%%%%%%%%%%%%%%%%%%%%%%%%%%%%%
\section{Introduction}
%%%%%%%%%%%%%%%%%%%%%%%%%%%%%%%%%%%%%%%%%%%%%%%%%%%%

Four dimensional $\cN=2$ gauged supergravity coupled to $n_v$-vector multiplets admits regular, static, BPS black holes which are asympotic to $AdS_4$. For very particular choices of prepotential, gauging parameters and background charges, there exists a remarkable analytic solution \cite{Cacciatori:2009iz} due to Cacciatori and Klemm (CK). These solutions can have spherical, flat or hyperbolic horizon geometries\footnote{The CK solutions generalize the original solution which has constant scalars and hyperbolic horizon geometry \cite{Caldarelli1999}. One can also couple $n_h$-hypermultiplets and there is also a rich solution space \cite{Halmagyi:2013sla} but in the current work we will restrict our focus to vector-multiplets. There has been some futher study of these CK solutions as well as related non-BPS $AdS_4$ black holes \cite{Bellucci2008b, Kimura:2010xe, Dall'Agata:2010gj, Hristov2011, Klemm:2012yg,Hristov:2013sya, Gnecchi:2012kb, Barisch:2011ui, Donos2012a, Donos:2012yi} } and can be interpreted in M-theory as M2-branes wrapped on Riemann surfaces \cite{Maldacena:2000mw, Gauntlett2002}.

To understand the structure of these new black holes we find it quite useful to compare with the much better understood, asymptotically flat, single-center, BPS black holes in $\cN=2$ ungauged supergravity \cite{Ferrara:1995ih, Behrndt:1997ny}. These asymptotically flat black holes preserve eight supercharges at infinity, four along the bulk of the solution and eight again at the horizon where the usual enhancement occurs. While the scalar fields are unfixed at infinity, they are typically fixed at the horizon giving rise to the moniker {\it attractor mechanism}.

For the black hole solutions of $\cN=2$ gauged supergravity relevant to the current work, the asymptotic $AdS_4$ region will preserve all eight supercharges, while along the bulk of the solution just two will be preserved and this is enhanced to four supercharges  at the horizon. The vector-multiplet scalars will tend to be fixed at infinity as well as at the horizon, albeit to different values; giving rise to a varying effective cosmological constant.

Whereas ungauged $\cN=2$ supergravity is {\it invariant} under an action of $Sp(2n_v+2,\RR)$, the gauged theory should be {\it covariant} under this action of the symplectic group. To enforce this covariance, one should work in a formalism which allows for magnetic gaugings. This is a somewhat complicated issue, the basic idea is that one can introduce magnetic gauge fields along with auxiliary tensor fields \cite{deWit:2005ub} but having said that, the explicit component Lagrangian of $\cN=2$ gauged supergravity with both magnetic and electric gaugings has not been written out. It should nonetheless be possible to extract the relevant formulae from \cite{deWit:2011gk}. In this work we follow the pragmatic approach of \cite{Dall'Agata:2010gj} and work with a theory which is a straightforward symplectic completion of \cite{Andrianopoli:1996vr}. 

Using this symplectically covariant formalism we solve the algebraic BPS equations for horizon geometries of the form $AdS_2\times \Sig_g$ with arbitrary electric and magnetic charges within a theory constructed from a general cubic prepotential. Schematically, the horizon equations express the charges and gauging parameters in terms of the scalar fields and metric components, a solution to this system constitutes inverting these and solving for the scalar fields and metric components in terms of the charges and gaugings. In ungauged supergravity, black hole horizons in theories with a general cubic prepotential were first systematically studied by Shmakova \cite{Shmakova:1996nz} where an implicit solution was provided up to set of $n_v$ real quadratic equations for $n_v$ variables. In addition, when $p^0=0$, an explicit solution was obtained. This has been extended in \cite{Ferrara:1996um, Ferrara:1997uz} \footnote{see also \cite{Pioline:2006ni} for a very nice derivation} where the explicit solution was found for situations where the special K\"ahler manifold is a symmetric space, the final answer being
\be
S_{BH}=\pi \sqrt{I_4(p^\Lam,q_\Lam)} 
\ee
where $I_4$ is the quartic invariant \eq{quartic}. 

The general solution obtained in the current work for black hole horizons in gauged $\cN=2$ supergravity is implicit up to a set of $n_v$-holomorphic quadratic equations in $n_v$-complex variables. Due to this holomorphicity, the solutions space is non-empty and at least zero-dimensional. These equations can also be explicitly solved when the special K\"ahler manifold is symmetric, giving rise to an analytic solution for the scalar fields and the entropy. We find that the entropy is related to, but not equal to, the quartic invariant $\cI_4$. Our solution for the entropy depends on both the charges and the gaugings. We also find an explicit solution when both one magnetic charge and one magnetic gauging parameter vanish 
\be
p^0=0,\ \ \ P^0=0\,.
\ee

It is of significant interest to show generally that these horizon solutions can be continued to the UV and connected to a BPS $AdS_4$ vacuum. There is currently no general analytic solution like the ungauged case, nonetheless we construct numerically a particular example of interest, that which corresponds to an internally spinning M2-brane wrapped on $\Sig_g$ in a Calabi-Yau fivefold.

This paper is organised as follows. In section 2. we introduce the ansatz and review the BPS equations for horizon solutions. We derive a form of the equations which is particularly amenable to solution and expand these equations for a general cubic prepotential. In section 3 we solve the BPS equations and  consider in more detail the solution for symmetric Special K\"ahler manifiolds. We present the canonical STU mode example which can be embedded in M-theory. We also numerically compute a particular dyonic black hole which corresponds to a spinning M2 brane wrapped on a Riemann surface in a Calabi-Yau fivefold.

%%%%%%%%%%%%%%%%%%%%%%%%%%%%%%%%%%%%%%%%%%%%%%%%%%%%
\section{The Algebraic Fixed Point Equations}
%%%%%%%%%%%%%%%%%%%%%%%%%%%%%%%%%%%%%%%%%%%%%%%%%%%%

We largely follow the conventions of \cite{Halmagyi:2013sla} but here we repeat the four dimensional black hole ansatz:
\bea\label{ansatz}
ds_4^2&=& e^{2U} dt^2- e^{-2U} dr^2- e^{2(V-U)} (d\tha^2+H(\tha)^2 d\vphi^2) \label{metAnsatz}\\
A^\Lam&=& \tq^\Lam(r) dt- p^\Lam(r) F'(\tha)  d\vphi \,, 
\eea
with
\be
H(\tha)=\left\{ \barr{ll} \sin \tha:& S^2\ (\kappa=1) \\ 
\sinh \tha:& \HH^2\ (\kappa=-1) \earr \right. 
\ee
and the scalar fields are radially dependant $z^i=z^i(r)$. This ansatz is general enough to describe asymptotically flat or $AdS_4$ black holes however this work we will primarily be interested in the horizon geometries where the radial dependance is fixed. As explained in \cite{Halmagyi:2013sla} one can easily include flat horizons.

The electric and magnetic charges are
\bea
p^\Lam &=& \frac{1}{4\pi } \int_{S^2} F^\Lam \label{elinv} \, , \\
 q_\Lam &\equiv& \frac{1}{4\pi} \int_{S^2} G_\Lam = -e^{2(V-U)} \cI_{\Lam\Sig} \tq'^\Sig +\cR_{\Lam\Sig} \kappa p^\Sig\, , 
  \label{maginv}
\eea
where $G_\Lam$ is the  symplectic-dual gauge field strength 
\be
G_{\Lam} \equiv \frac{\delta \cL}{ \delta F^\Lam}=R_{\Lam \Sig} F^\Lam -\cI_{\Lam \Sig} *F^\Sig \, . 
\ee
This ansatz works in theories with just electric gaugings, we develop below the generalization needed to include magnetic gaugings.

%%%%%%%%%%%%%%%%%%%%%%%%%%%%%%%%%%%%%%%%%%%%%%%%%%%%
\subsection{$AdS_2\times \Sig_g$ Fixed Point Equations}
%%%%%%%%%%%%%%%%%%%%%%%%%%%%%%%%%%%%%%%%%%%%%%%%%%%%

The horizon of a static BPS black hole is of the form $AdS_2\times \Sig_g$. This requires  the metric functions to take the form
\be
e^U=\frac{r}{R_1}\,,\ \ \ \ \ e^V=\frac{rR_2}{R_1}
\ee
and all scalar fields are constants
\be
z^i(r)=z_0^i\,.
\ee
When the theory has just electric gaugings, the horizon equations are \cite{Cacciatori:2009iz, Halmagyi:2013sla, Dall'Agata:2010gj}
\bea
p^\Lam Q_\Lam&=& 1 \\
e^{-i\psi} L^\Lam Q_\Lam&=& \frac{i}{2R_1} \\
\cZ&=& e^{i\psi}\frac{R_2^2}{2R_1} \\
\kappa p^\Lam&=& -\frac{2R_2^2}{R_1} \Im (e^{-i\psi} L^\Lam) - R_2^2 \cI^{\Lam\Sig} Q_\Sig \\
q_\Lam &=& -\frac{2R_2^2}{R_1} \Im( e^{-i\psi}\cM_{\Lam}) - R_2^2 \cR_{\Lam \Sig} \cI^{\Sig \Delta} Q_\Delta \,.
\eea
We have introduced a certain phase $e^{i\psi}$ of the supersymmetry parameter 
\be
\eps_A= e^{U/2} e^{i\psi} \eps_{0A}
\ee
where $\eps_{0A}$ is an $SU(2)$ doublet of constant spinosr satsifying the two projections
\bea
\eps_{0A}&=&  i\eps_{AB} \gam^0 \eps_0^B\,, \\
\eps_{0A}&=&  - (\sig^3)_A^{\ B} \gam^{01} \eps_{0B}\,.
\eea

%%%%%%%%%%%%%%%%%%%%%%%%%%%%%%%%%%%%%%%%%%%%%%%%%%%%
\subsection{Symplectic Invariant Equations}
%%%%%%%%%%%%%%%%%%%%%%%%%%%%%%%%%%%%%%%%%%%%%%%%%%%%

Working with magnetic and electric gaugings requires introducing a fair amount of data from special geometry and we have included the necessary background in appendix \ref{app:special}. A key result is that
we can expand any symplectic vector in terms of $(\cV,U_i)$ and their conjugates. For example the background charges $\cQ$ and gaugings\footnote{In trying to formulate a sensible symplectic covariant notation we have slightly altered the standard notation and denoted by $Q_\Lam$ what is typically denoted $P_\Lam$.  Furthermore this is different still from the notation in \cite{Dall'Agata:2010gj} where the gauging parameters were denoted $\cG=\bpm g^\Lam \\ g_\Lam \epm$ and $\cL=\langle \cG,\cV\rangle$. We hope the interested reader will persevere regardless.}
\be
\cQ= \bpm \kappa p^\Lam \\ q_\Lam \epm\,,\ \ \ \ \ \cP = \bpm P^\Lam \\ Q_\Lam \epm
\ee
can be expanded as
\bea
\cQ&=& i\cZbar \cV-i \cZ \cVbar+i \cZ^{\ibar} \Ubar_{\ibar}- i \cZbar^{i} U_{i} \\
\cP&=& i\cWbar \cV-i \cW \cVbar+i \cW^{\ibar} \Ubar_{\ibar}- i \cWbar^{i} U_{i} 
\eea
so that
\be
\cZ=\langle \cQ,\cV  \rangle\,,\ \ \ \ \ \cW=\langle \cP,\cV  \rangle\,, \ \ \ \ \  \cZ_i=\langle \cQ,U_i  \rangle\,,\ \ \ \ \ \cW_i=\langle \cP,U_i  \rangle\,.
\ee
Key objects for our analysis are the symplectic matrices $\cM$ and $\Om$ (see appendix \ref{app:special}) which together satisfy
\bea
\Om\cM\cV=-i\cV\,,\ \ \ \ \Om\cM U_i = i U_i \,.
\eea
Using these one arrives at the symplectically covariant horizon equations \cite{Dall'Agata:2010gj}
\bea
\cQ - R_2^2 \Om \cM \cP&=&-4 \, \Im(\cZbar \cV) \label{symphor1} \\
\cZ &=&  e^{i\psi} \frac{R_2^2}{2R_1}\label{symphor2}  \\
 \langle \cP,\cQ \rangle &=& 1\,.\label{PQ1}
\eea
Since the gravitino is charged, Dirac quantization implies
\be
 \langle \cP,\cQ \rangle\in \ZZ
\ee
while the BPS condition \eq{PQ1} selects this integer to be unity.

Eq \eq{symphor1} is itself a symplectic vector thus we can extract its components by contracting \eq{symphor1} with $\cV$ and $U_i$ resulting in
\bea
\cW&=&\frac{i}{R_2^2} \cZ \,, \label{WZ1}\\
\cW_i&=&\frac{i}{ R_2^2} \cZ_i\,.\label{WZ2} 
\eea
It is worth noting that \eq{symphor1}-\eq{PQ1} constitute $2n_v+5$ real equations for the $2n_v+3$ variables $\{z^i,R_1,R_2,\psi\}$ so generically we expect there to be two constraints on the solution space which is parameterized by $(p^\Lam,q_\Lam)$.

In the limit  $\cP\ra 0$ of vanishing gauging parameters, these equations reduce to the attractor equations of ungauged supergravity, apart from \eq{PQ1} which should not be enforced in that limit.  One qualitative difference in the gauged case is that the LHS of \eq{symphor1} depends on the scalar fields through $\Om\cM \cP$ term which makes it somewhat more complicated to disentangle these equations . The central goal of this paper is to decouple these equations such that the scalar fields and radii are solved for in terms of the charges $\cQ$ and the gaugings $\cP$.

Using \eq{WZ1} and \eq{WZ2} we get
\be
iR_2^2 \cP= i\cZbar \cV+i \cZ \cVbar-  i\cZ^{\ibar} \Ubar_{\ibar}-i\cZbar^{i} U_{i}
\ee
and so  \eq{symphor1}-\eq{PQ1} are equivalent to 
\be\fbox{$
\cQ+ iR_2^2\cP = 2i \cZbar \cV - 2i \cZbar^i U_i \label{QiP}
$}\,.
\ee
The derivation of \eq{QiP} is a key step in this analysis as we have eliminated the dependance of the scalar fields on the LHS in \eq{QiP}, nonetheless there remains a dependance on $R_2$.  Defining the holomorphic quantities
\bea
\fp^\Lam &=& \kappa p^\Lam + i R_2^2 P^\Lam\,, \label{pdef}\\
\fq_\Lam&=& q_\Lam + i R_2^2 Q_\Lam \label{qdef}
\eea
and using details from Appendix \ref{app:special}. we can expand \eq{QiP} explicitly for an arbitrary cubic prepotential:
\bea
\fp^0&=& 2i e^{K/2} \bslb \cZbar +12 i e^K D_{y,i}\cZbar^i\bsrb \label{fpfq1} \\
\fp^i &=& \fp^0 z^i -2i e^{K/2} \cZbar^i \label{fpfq2}\\
\fq_0 &=& -\fp^0 D_z +6i e^{K/2} \cZbar^i D_{z,i} \label{fpfq3} \\
\fq_i &=& \bslb 3 \fp^j-6ie^{K/2} \cZbar^j \bsrb D_{z,ij} \label{fpfq4}\,.
\eea
This system is $2n_v+2$ holomorphic equations for the $2n_v+1$ complex variables $\{\cZ,\cZ^i,z^i \}$. This
must be supplemented with \eq{PQ1} and in addition we must solve for $R_2$. Like \eq{symphor1}-\eq{PQ1} this generically this gives a system with 2 additional real constraints on $(\cP,\cQ)$.

It is interesting that the variables which we formulate the equations in terms of, namely $(\fp^\Lam,\fq_\Lam)$ are almost objects which should be {\it specified} before solving any given background. In particular $(P^\Lam,Q_\Lam)$ define the theory under consideration while $(p^\Lam,q_\Lam)$ define the vacuum selection sector, the glitch is of course that $(\fp^\Lam,\fq_\Lam)$ also involve $R_2$. Our strategy is to solve for all variables in terms of $(\fp^\Lam,\fq_\Lam)$ and then use our final equation to solve for $R_2$ in terms of the gaugings and the charges.

%%%%%%%%%%%%%%%%%%%%%%%%%%%%%%%%%%%%%%%%%%%%%%%%%%%%
\section{Solving the Horizon Equations}
%%%%%%%%%%%%%%%%%%%%%%%%%%%%%%%%%%%%%%%%%%%%%%%%%%%%

We now solve the horizon equations \eq{fpfq1}-\eq{fpfq4} for models derived from a general cubic prepotential 
\be
F=D_{ijk} \frac{X^i X^j X^k}{X^0}
\ee
and with arbitrary charges $\cQ$ and gaugings $\cP$. To this end, we construct using \eq{fpfq1}-\eq{fpfq4} a set of $n_v$ holomorphic, quadratic equations:
\bea
D_{ijk}\fp^j \fp^k  &=& \frac{1}{3} \fp^0 \fq_i -4 e^K D_{ijk}\cZbar^j \cZbar^k \label{ZiZj}\,.
\eea
We define 
\bea
\Pi_i&=& D_{ijk} \fp^j \fp^k -\frac{1}{3} \fp^0 \fq_i\,,  \label{PiDef}\\
\hcZbar^i&=&2i e^{K/2} \cZbar^i 
\eea
so that \eq{ZiZj} becomes
\be \fbox{$
\Pi_i=D_{ijk}\hcZbar^j \hcZbar^k
$}\label{PiEq}\,.\ee

%%%%%%%%%%%%%%%%%%%%%%%%%%%%%%%%%%%%%%%%%%%%%%%%%%%%
\subsection{$\fp^0= 0$}
%%%%%%%%%%%%%%%%%%%%%%%%%%%%%%%%%%%%%%%%%%%%%%%%%%%%

When $\fp^0=0$ we can explicitly solve \eq{fpfq2}:
\be
e^{K/2}\cZbar^i= \frac{i \fp^i}{2}
\ee
and then from \eq{fpfq4} we solve for the scalar fields $z^i$
\be
z^i=\frac{1}{6} (D_{\fp}^{-1})^{ij} \fq_j
\ee
where 
\be
(D_\fp^{-1})^{ij}D_{\fp,jk}=\delta^{i}_k\,.
\ee
We then obtain $(R_1,\psi)$ from \eq{fpfq1}:
\be
\cZbar=6 e^{K/2} D_{y,i} \fp^i \,.
\ee
Finally we have the relation
\be
0=2 \fq_0 +\frac{1}{6}  (D_{\fp}^{-1})^{ij} \fq_i \fq_j
\ee
from which one can obtain $R_2^2$ in terms of $(p^
\Lam,q_\Lam,P^\Lam,Q_\Lam)$ as well as one constraint on the charges.

%%%%%%%%%%%%%%%%%%%%%%%%%%%%%%%%%%%%%%%%%%%%%%%%%%%%
\subsection{$\fp^0\neq 0$}
%%%%%%%%%%%%%%%%%%%%%%%%%%%%%%%%%%%%%%%%%%%%%%%%%%%%
Assuming more generally that $\fp^0\neq 0$ we can then solve \eq{fpfq2} for $z^i$
\be
z^i= \frac{\fp^i + \hcZbar^i}{\fp^0}\,. \label{ziSol2}
\ee
Then \eq{fpfq1} can be used to solve for $\cZ$ and thus:
\bea
e^{i\psi} \frac{R_2^2}{R_1}&=& \frac{\fp^0}{2ie^{K/2}}-6 e^{K/2} D_{y,i} \hcZbar^i \label{ZSol}
%&=& \sqrt{2}\fp^0 D_y^{1/2}+\frac{3i}{\sqrt{2}} \frac{ D_{y,i} \hcZbar^i }{D_y^{1/2}}\\
%&=&  \sqrt{2} D_y^{1/2}\Bslb \fp^0+2i g_{ij} y^i \hcZbar^j \Bsrb
\eea
From \eq{ZSol} we can extract the solution for one radius $R_1$ and the phase of the spinor $\psi$ as a function of the the charges, gaugings and the second radius (which governs the entropy) $(\cP,\cQ,R_2)$.
In principle once the solution for $\hcZbar^i$ has been obtained one immediately obtains $z^i$ from \eq{ziSol2}. The final step to obtain $R_2$ is to solve  \eq{fpfq3}, which one can show is equivalent to
\bea
%\fq_0 &=& -\fp^0 D_z +3\hcZbar^i D_{z,i} \label{fpfq3} \\
%(\fp^0)^2 \fq_0 &=&  -D_{\fp} + (3\fp^i +2 \hcZbar^i )\Pi_i \label{q0Eq} \\
\fp^0 \cI_{2}(\fp,\fq) &=&   \hcZbar^i \Pi_i \label{q0Eq}
\eea
where we have introduced the quadratic symplectic invariant
\be
\cI_2(\fp,\fq)= \frac{D_{\fp}}{\fp^0} - \frac{1}{2} \fp^\Lam \fq_\Lam\,. \label{quadInv} 
\ee
This gives equations from both the real and imaginary parts, one of which can be used to solve for $R_2$ and the other gives a constraint on the charges. To extract a completely explicit solution  to \eq{q0Eq}, we will restrict to the case of $\cM_v$ being a symmetric space. Finally one must also impose the Dirac quantization condition \eq{PQ1}.

It is important to check regularity of any resulting solution. There is an important constraint on the scalar fields that the metric $g_{i\jbar}$ be finite and one must have $R_2^2 >0$ so that the spacetime metric is well defined. From \eq{ZSol}  we see that $R_1$ can always be taken real since any phase is absorbed into $e^{i\psi}$.

%%%%%%%%%%%%%%%%%%%%%%%%%%%%%%%%%%%%%%%%%%%%%%%%%%%%
\subsection{Symmetric Spaces}
%%%%%%%%%%%%%%%%%%%%%%%%%%%%%%%%%%%%%%%%%%%%%%%%%%%%

When $\cM_v$ is a symmetric space, we can solve \eq{PiEq} explicitly. The key object is the contravariant three-tensor \cite{Gunaydin:1983bi, deWit:1992wf}
\bea
\hD^{ijk} = \frac{g^{il}g^{jm}g^{kn}D_{ijk}}{D_y^2}
\eea
which for symmetric spaces is constant and satisfies
\bea
\hD^{ijk}D_{j(lm}D_{np)k}&=&\frac{64}{27}\delta^i_{(l}D_{mnp)}\,, \\
D_{ijk}\hD^{j(lm}\hD^{np)k}&=&\frac{64 }{27}\delta_i^{(l}\hD^{mnp)}\,.
\eea
The solution to \eq{PiEq} is then
\bea
\hcZbar^i&=&\pm \sqrt{\frac{27}{64}}\frac{  \hD^{ijk}\Pi_j\Pi_k}{ \sqrt{\hD_{\Pi}}} \label{ZiSol}
\eea
where we have defined
\bea
\hD_{\Pi}&\equiv& \hD^{ijk} \Pi_i \Pi_j \Pi_k \non \\
&=& \frac{16}{27}(\fp^0)^2 \bslb  \cI_{4}(\fp,\fq)+4 \cI_2(\fp,\fq)^2\bsrb
\eea
and we have introduced the familiar symplectic invariant
\bea
\cI_4 (\fp ,\fq)&=& - \blp \fp^\Lam \fq_\Lam \brp^2-\frac{1}{16} \fp^0 \hD^{ijk}  \fq_i \fq_j \fq_k+4\fq_0 D_{ijk}  \fp^i \fp^j \fp^k +\frac{9}{16}D_{ijk} \hD^{ilm} \fp^j \fp^k \fq_l \fq_m \,. \label{quartic}
\eea
We are now in a position to unpack \eq{q0Eq}
and we get
\be
2 \cI_2(\fp,\fq)=\pm \sqrt{\cI_4(\fp,\fq)+4\cI_{2}(\fp,\fq)^2}\,,
\ee
the only regular solution takes the $+$-sign in \eq{ZiSol}:
\bea
0=\cI_4(\fp,\fq)\,. \label{quarticzero}
\eea

Recalling the definitions \eq{pdef} and \eq{qdef}, then separating \eq{quarticzero} into real and imaginary parts gives two polynomials in $R_2^4$
\bea
0&=& a_0+ a_4 R_2^4 + a_8 R_2^8   \label{R2Eq1}\\
0&=& a_2 R_2^2+ a_6 R_2^6   \label{R2Eq2}
\eea
where the $a_i$ are quartic invariants of the diagonal action on the charges and gaugings. The explicit expressions for the $a_i$ can easily be obtained from \eq{quartic} but do not appear to be particularly enlightening for our purposes. Having said that, we do have
\be
a_0=\cI_4(\kappa p,q)\,,\ \ \ \ a_{8}=\cI_{4}(P,Q)\,.
\ee

If either $a_2=0$ or $a_6=0$ then for regularity, both must vanish. In this case, we have
\be
R_2^4=\frac{-a_4\pm \sqrt{a_4^2-a_0a_8}}{2a_8}\,. \label{R2quad}
\ee
and no extra constraint. Regularity implies that $R_2^4>0$ which gives a bound on the space of charges.

Otherwise, if both $a_2\neq 0$ and $a_6\neq 0$ we have
\be
R_2^4=-\frac{a_2}{a_6} \label{R2Sola2a6}
\ee
and the constraint
\be
0=a_0a_6^2-a_2a_4a_6+a_2^2 a_8\,. \label{chargeconstraint}
\ee
Again we have the bound $R_2^4>0$.

One can also use this explicit solution to obtain an explicit expression for $\cZ$ \eq{ZSol} and thus $R_1$. It is of some interest to express this in a compact form in terms of the explicit invariants of the diagonal symplectic action on $(\cQ,\cP)$ \cite{NickWip}.

%%%%%%%%%%%%%%%%%%%%%%%%%%%%%%%%%%%%%%%%%%%%%%%%%%%%
\subsubsection{Revisiting Black Hole Horizons in Ungauged $\cN=2$ Supergravity}
%%%%%%%%%%%%%%%%%%%%%%%%%%%%%%%%%%%%%%%%%%%%%%%%%%%%

While it is interesting to consider the limit $\cP\ra 0$ of the above solution to connect with BPS black hole horizons in ungauged supergravity, this limit is not regular. Nonetheless we note that the key equations in \cite{Shmakova:1996nz} are identical in form to \eq{PiEq} but real instead of holomorphic. In that work the entropy is given by
\bea
S&=&\frac{\pi }{3p^0}\sqrt{\frac{4}{3} (\Delta_i \tx^i)^2-9\blp p^0  p^\Lam q_\Lam -2 D_{ijk} p^i p^j p^k \brp^2} 
\eea
where
\bea
\Delta_i&=& 3 D_{ijk} p^j p^k -p^0 q_i
\eea
is essentially identical to $\Pi_i$ in \eq{PiDef} and 
\be
\tx^i =\sqrt{12} e^{K/2} |\cZ| y^i
\ee
solve the set of $n_v$ real quadratic equations
\be
\Delta_i= D_{ijk} \tx^j\tx^k\,. \label{DeltaEq}
\ee
We note that similarly as above, when $\cM_v$ is a symmetric space one can solve \eq{DeltaEq} with
\be
\tx^i= \frac{3\sqrt{3}}{8}\frac{\hD^{ijk} \Delta_i \Delta_k}{\sqrt{\hD^{ijk} \Delta_i \Delta_j \Delta_k}}
\ee
from which we find 
\be
S=\pi \sqrt{I_4(p,q)}\,.
\ee
This provides an alternative derivation from that in \cite{Pioline:2006ni} for the explicit entropy of BPS black holes when the special K\"ahler manifold is a symmetric space.

%%%%%%%%%%%%%%%%%%%%%%%%%%%%%%%%%%%%%%%%%%%%%%%%%%%%
\subsection{Embedding Dyonic Black Holes in M-theory} \label{sec:Mtheory}
%%%%%%%%%%%%%%%%%%%%%%%%%%%%%%%%%%%%%%%%%%%%%%%%%%%%

The model with $n_v=3$ and prepotential given by
\bea
F&=& -\frac{X^1 X^2 X^3}{X^0}  
\eea
is commonly refered to as the STU-model. In gauged supergravity this model has considerable complexity due to the gaugings $(P^\Lam,Q_\Lam)$ and the spectrum of solutions within this model varies discontinuously with these gauging parameters. 

For a particular set of gaugings, there is a known embedding of this gauged supergravity theory into M-theory on $S^7$ \cite{Duff:1999gh, Cvetic1999b}. This model is given by
\bea
P^\Lam&=& -(0,g,g,g)\,,\ \ \  Q_\Lam=(g,0,0,0) \,.
\eea
If we rotate to another duality frame using 
\bea
\cS&=& \bpm A& B \\ C & D\epm\,,\ \ \ 
A=D= \diag \{1,0,0,0\}\,,\ \ \ 
B=-C= \diag\{0,1,1,1\}
\eea
we get 
\bea
F&=& -2i \sqrt{X^0 X^1 X^2 X^3} \\
P^\Lam &=&0\,, \ \ \  \ Q_{\Lam}= g\,.
\eea
and so the gaugings in this frame are purely electric. There exists a remarkable analytic solution \cite{Cacciatori:2009iz}  in this duality frame for black holes which are {\it purely magnetic}. The horizon geometries which arise as the IR of these magnetic black holes are specified by three independant charges; there is four charges and just one constraint \eq{PQ1}, $a_2=a_6=0$ and the radius $R_2$ comes from \eq{R2quad}.

 More generally, from the computations above, we see that in this model the general space of solutions is parameterized by {\it six} charges; the eight charges have two constraints, one from \eq{PQ1} and the extra constraint \eq{chargeconstraint}. So for these dyonic configurations the entropy comes from \eq{R2Sola2a6}. One of course needs to check that $a_2/a_6<0$ but one finds that this can generically be arranged.

\begin{figure}[bth!]
 \begin{center}
\vspace{0pt}
\includegraphics[scale=.6]{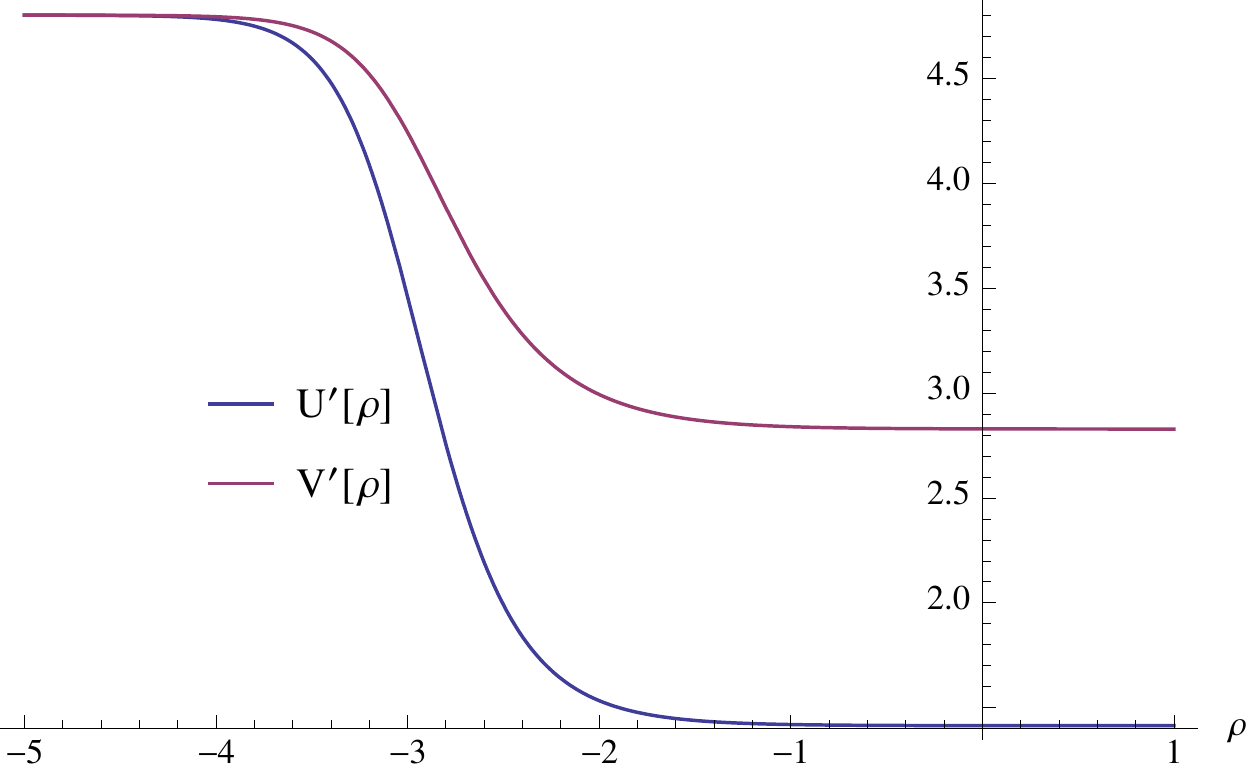}
\hspace{.5cm}
\vspace{1em}
\caption {The metric functions $U'(\rho)$ and $V'(\rho)$}
\label{fig:flow3vecsMetric}
\end{center}
\end{figure}

\begin{figure}[bth!]
 \begin{center}
\vspace{0pt}
\includegraphics[scale=.6]{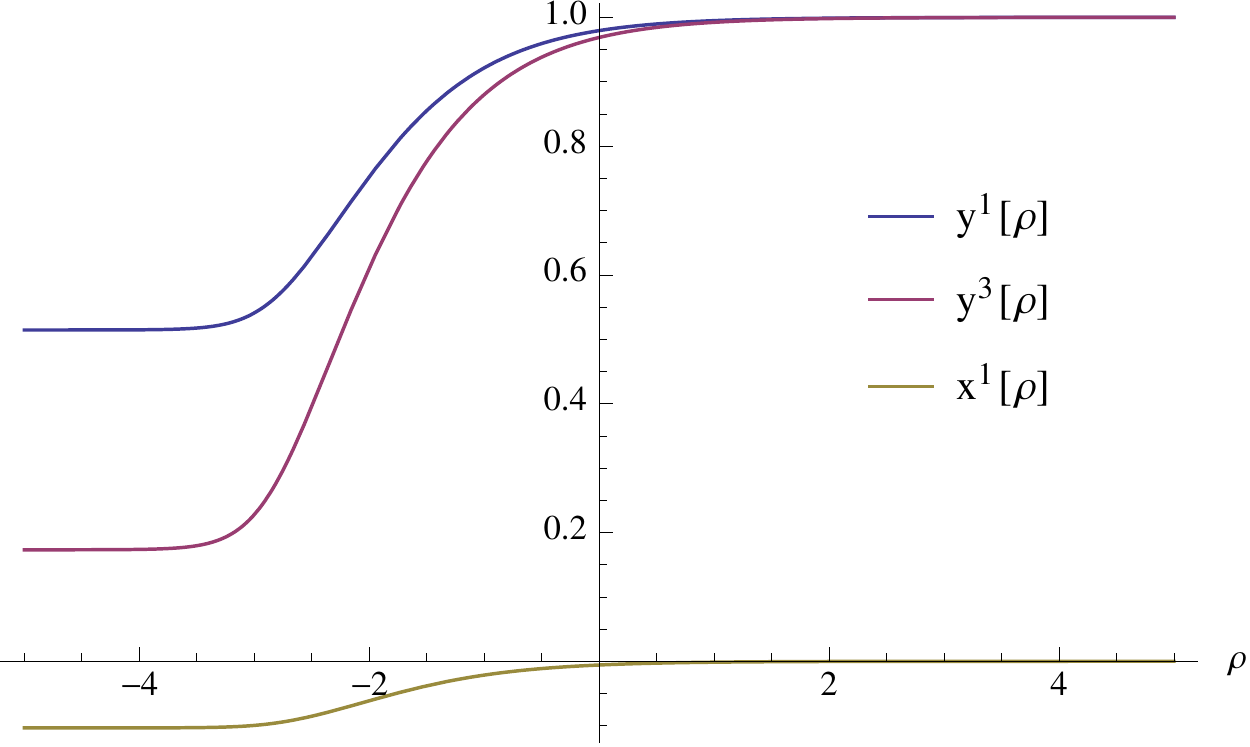}
\hspace{.5cm}
\vspace{1em}
\caption {The scalar functions $y^1=y^2$, $y^3$ and $x^1$ }
\label{fig:flow3vecsScalars}
\end{center}
\end{figure}

While there is a nice analytic solution for the entire magnetic black hole in this duality frame, we have been unable to find an analytic solution for the black hole with additional electric charges. With non-trivial electric charges, the real part of the vector multiplet scalars $x^i$ is turned on, which complicates significantly the formulae. Nonetheless we have computed numerically the entire black hole solution for a particular choice of electric and magnetic charges and $\Sig_g=S^2$. The plots are shown in fig. \ref{fig:flow3vecsMetric} and \ref{fig:flow3vecsScalars} for the radial variable $\rho=e^{U_{IR}}\log r$. The charges have been chosen to be \be
p^\Lam=(5,-1,-1,-2),\ \ \ q_\Lam=(0,1,-1,0)
\ee
With these choices of charges, we find that $y^1=y^2$, $x^1=-x^2$ and $x^3=0$ along the whole flow although one does not need to restrict to such a simple dyonic configuration.

\vskip 3cm
These black holes have a simple interpretation in terms of wrapped branes \cite{Maldacena:2000mw}. Consider the local Calabi-Yau five fold $X_5$ which is the product of four line-bundles over a Riemann surface $\Sig_g$:
\be
\xymatrix{
\oplus_{\Lam=0}^{3}\cL_{p^\Lam}\ar[r] & X_5 \ar[d]\\
&  \Sig_g\\} \non
\ee
The four magnetic charges of the gauged supergravity are related to the Chern-numbers of these line bundles and the BPS condition \eq{PQ1} is tantamount to $c_1(X_5)=0$. The additional electric charges which we have found in this work correspond to spin of these M2-branes along the particular $U(1)$ isometries of $X_5$. 

%%%%%%%%%%%%%%%%%%%%%%%%%%%%%%%%%%%%%%%%%%%%%%%%%%%%
\section{Conclusions}
%%%%%%%%%%%%%%%%%%%%%%%%%%%%%%%%%%%%%%%%%%%%%%%%%%%%

We have solved the horizon equations for BPS black hole in $\cN=2$ gauged supergravity with $n_v$-vector multiplets by employing a generalization of the types of special geometry techniques that have proved useful in the past for the study of BPS black holes in ungauged supergravity. The level of detail of our solution is essentially equivalent to the state of the art for ungauged supergravity but it is curious that our set of quadratic equations \eq{PiEq} are holomorphic whereas the similar equations in the ungauged case are real. This  holomorphicity guarantees a non-vanishing solution space but one must still check regularity. It would seem that a reasonable strategy to understand black holes in gauged supergravity would be to continue to utilize the many tools which have been put to good use in the ungauged theory, much as we have done in this paper. The study of higher derivative corrections in the gauged case presents a formidable challenge.

By the general principles of holography, it is desirable to embed these black holes into a UV complete theory of gravity such as string or M-theory, but currently the only known embedding is for the particular STU-model discussed in section \ref{sec:Mtheory}. While it would certainly be a welcome development to derive quite general models of gauged supergravity with vector multiplets from M-theory, in lieu of that, it is natural to include hypermultiplets where there is a richer set of known embeddings (one can find numerous explicit examples in \cite{Gauntlett:2009zw, Donos:2010ax, Cassani:2012pj}). Studying M-theory enbeddings might help understand the physical meaning behind the combinations $(\fp^\Lam,\fq_\Lam)$. These are a combination of charges, which specify the vacuum selection sector and gauging parameters which appear in the action and define the theory. The pairing does not appear to be so natural, it would be nice to have a physical understanding of why the BPS equations have a neat form when this combination are used.

In this work we have largely focussed on the horizon geometries but it would be quite helpful to have a general proof that such horizons can be UV completed to an $\cN=2$, $AdS_4$ solution. In the case of the CK black holes where an analytic solution is at hand, one can easily check that the range of magnetic charges for which a regular black hole exists is identical to the range of charges for which a horizon geometry exists. One can thus conclude that the CK solution space exhausts the space of magnetic black holes in that particular gauged supergravity theory. It would be a significant development to have analytic solutions for more general dyonic black holes much like the general solution is known in ungauged supergravity. We hope to return to these and other issues in the immediate future.

%%%%%%%%%%%%%%%%%%%%%%%%%%%%%%%%%%%%%%%%%%%%%%%%%%%%
\vskip 5mm
\noindent {\bf Acknowledgements:} It is a pleasure to thank Mohab Abou-Zeid, Davide Cassani, Atish Dabholkar, Sheer El-Showk, Boris Pioline, Martin Ro$\rm{\check{c}}$ek and Brookie Phipps Williams for useful conversations and communcation. I would also like to acknowledge the collaboration with Michela Petrini and Alberto Zaffaroni on related topics.

%%%%%%%%%%%%%%%%%%%%%%%%%%%%%%%%%%%%%%%%%%%%%%%%%%%%
\begin{appendix}
%%%%%%%%%%%%%%%%%%%%%%%%%%%%%%%%%%%%%%%%%%%%%%%%%%%%

%%%%%%%%%%%%%%%%%%%%%%%%%%%%%%%%%%%%%%%%%%%%%%%%%%%%
\section{Some Special Geometry}\label{app:special}
%%%%%%%%%%%%%%%%%%%%%%%%%%%%%%%%%%%%%%%%%%%%%%%%%%%%

In this appendix we collect various formula for the special geometry of $\cM_v$ the vector-multiplet scalar manifold, essentially to establish a consistent notation. All this material is well known but there are numerous different conventions in the literature. We have the cubic prepotential
\bea
F=D_{ijk}\frac{X^i X^jX^k}{X^0}\,, \label{cubicPrepot}
\eea
where $i=1,\ldots , n_v$ and will often work with special co-ordinates
\bea
X^\Lam&=& (1,z^i)\,, \\
z^i&=& x^i +i y^i\,,
\eea
where $\Lam=0,\ldots , n_v$. We use the notation
\be
D_{z}= D_{ijk} z^i z^j z^k\,, \ \ \ D_{z,i} = D_{ijk} z^j z^k \,,\ \  \ D_{z,ij}=D_{ijk} z^k
\ee
and similarly for $(D_y,D_{y,i},D_{y,ij})$ etc.
 The metric on $\cM_v$ can be explicitly given in terms of $D_{ijk}$:
 \bea
g_{ij}&=&\frac{\del K}{\del \tau^i \del \taubar^{\jbar}}\non \\
&=& -\frac{3D_{y,ij}}{2D_y}+\frac{9D_{y,i}D_{y,j}}{4D_y^2} \\
g^{ij}&=& -\frac{2}{3}D_y(D_y^{-1})^{ij}+2 y^i y^j
\eea
where $K$ is the K\"ahler potential and
\be 
e^{-K}=-8 D_y\,.
\ee

The complex scalar fields $z^i$ of the vector multiplets are co-ordinates on $\cM_v$ but special geometry requires consideration of certain sections of an $Sp(2n_v+2,\RR)$ bundle over $\cM_v$. These sections are denoted $(X^\Lam, F_\Lam)$ however the more natural objects are the rescaled sections:
\be
\cV= \bpm  L^\Lam \\ M_\Lam \epm =e^{K/2} \bpm  X^\Lam \\ F_\Lam \epm
\ee
which satisfy
\bea
M_\Lam &=&  \cN_{\Lam \Sig} L^\Sig \\
\cN_{\Lam \Sig}&=& \cR_{\Lam \Sig} + i\, \cI_{\Lam \Sig}\,.
\eea
One should note that $\cR_{\Lam \Sig}$ and $\cI_{\Lam \Sig}$ are the gauge kinetic and topological terms in the $\cN=2$ Lagrangian. An explicit expression for $\cN_{\Lam \Sig}$ is
\be
\cN_{\Lam \Sig} = \Fbar_{\Lam \Sig}+2i \frac{\Im F_{\Lam \Delta} \Im F_{\Sig \Upsilon } X^{\Delta} X^{\Upsilon}} {\Im F_{\Delta \Upsilon} X^{\Delta} X^{\Upsilon}}
\ee  
where
\be
F_{\Lam \Sig}= \del_\Lam \del_{\Sig}F\,.
\ee

We also use the covariant derivative of the sections which are defined to be
\be
U_i \equiv  D_i \cV= \del_i \cV + \half (\del_i K) \cV\,.
\ee
The components of $U_i$ are
\be
U_i = \bpm  f^i_\Lam \\ h_{i\, \Lam}\epm
\ee
and satisfy
\be
h_{i\, \Lam} = \cNbar_{\Lam \Sig} f_i^{\Sig}\,.
\ee
Of particular importance is the symplectic inner product $\langle.,. \rangle$
\be
\langle A,B\rangle = B^\Lam A_\Lam - A^\Lam B_\Lam
\ee 
under which we have
\bea
\langle \cV,\cVbar \rangle &=& -i\,,\ \ \ \ \langle U_i,\Ubar_{\jbar}\rangle = ig_{i\jbar} \,.
\eea
This allows us to expand a symplectic section in terms of $(\cV,U_i)$ and their conjugates. For example
 the background charges are given by
\bea
 \cQ&=& i\cZbar \cV-i \cZ \cVbar+i \cZ^{\ibar} \Ubar_{\ibar}- i \cZbar^{i} U_{i}
\eea
so that
\be
\cZ=\langle \cQ,\cV  \rangle\,,\ \ \ \ \  \cZ_i=\langle \cQ,U_i  \rangle\,.
\ee

Using \eq{cubicPrepot} we can compute the components of $\cV$:
\be
\cV=\bpm L^\Lam \\ M_{\Lam} \epm = e^{K/2} \bpm X^\Lam \\ F_\Lam \epm \ \ {\rm where}\ \  
\left\{ \barr{rcl}X^{0}&=&1 \\ X^i &=&z^i\\ F_0&=&-D_z \\ F_i &=& 3 D_{z,i} \earr \right.
\ee
The components of $U_i$ are 
\bea
(U_i)^0 &=& D_i L^0 = -12 i e^K D_{y,i}e^{K/2} \\
(U_i)^j &=&D_i L^j= e^{K/2} (\delta^j_i -12i e^K D_{y,i} z^j ) \\
(U_i)_0 &=& D_i M_0= e^{K/2}\bslb  -3D_{z,i}+12i e^K D_{y,i}D_z  \bsrb \\
(U_i)_j &=&D_i M_j =  e^{K/2}\bslb  6D_{z,ij}-36i e^K D_{y,i}D_{z,j}  \bsrb
\eea

Further objects we use in the main computation are
\bea
\cM&=&\bpm 1 &  -\cR \\0  &1 \epm \bpm \cI & 0 \\ 0& \cI^{-1} \epm\bpm 1 &  0 \\ -\cR  &1 \epm =\bpm A & B \\ C & D\epm \non 
\eea
where
\bea
A&=& \cI+\cR\cI^{-1} \cR \non  \\
D&=& \cI^{-1} \non \\
B&=& C^T= -\cR \cI^{-1} \non 
\eea
combined with the symplectic form 
\be
\Om=\bpm 0 &-1\!\!1\\ 1\!\!1 &0\epm\,.
\ee
Using these explicit expressions one can observe the property
\be
\Om \cM \cV=-i \cV\,,\ \ \ \Om \cM U_i = iU_i\,.
\ee

\end{appendix}

%%%%%%%%%%%%%%%%%%%%%%%%%%%%%%%%%%%%%%%%%%%%%%%%%%%%

\providecommand{\href}[2]{#2}\begingroup\raggedright\endgroup

\end{document}